# Dynamic Cluster Head Selection Using Fuzzy Logic on Cloud in Wireless Sensor Networks

Payal Pahwa, Deepali Virmani, Akshay Kumar, Sahil, Vikas Rathi, Sunil Swami

*Department of Computer Science and Engineering/Information Technology*
*Bhagwan Parshuram Institute of Technology, Sector 17, Rohini, New Delhi – 110089, India*

**Abstract**

One of the most vital activities to reduce energy consumption in wireless sensor networks is clustering. In clustering, one node from a group of nodes is selected to be a cluster head, which handles majority of the computation and processing for the nodes in the cluster. This paper proposes an algorithm for fuzzy based dynamic cluster head selection on cloud in wireless sensor networks. The proposed algorithm calculates a Potential value for each node and selects cluster heads with high potential. The proposed algorithm minimizes cluster overlapping by spatial distribution of cluster heads and discards malicious nodes i.e. never allows malicious nodes to be cluster heads.

© 2014 The Authors. Published by Elsevier B.V.

Selection and peer-review under responsibility of scientific committee of Missouri University of Science and Technology.

## 1. Introduction

Owing to the rampant development of communication, computation and hardware technology in the past decade, Wireless Sensor Networks (WSNs) have emerged in the fields of agriculture, homeland security, navy, weather forecasting etc. [1]. WSNs consist of spatially distributed sensor nodes which communicate wirelessly, have limited energy and computation power. A huge amount of research on WSNs is directed towards increasing the network lifetime of WSNs by making them energy efficient. One of the key activities to increase network lifetime in WSNs is clustering.

Clustering in WSNs is the process of dividing the WSN into smaller groups, aimed to increase network lifetime. Each group/cluster consists of a cluster head/ sink node. The cluster head is responsible for majority of the processing and computation of the cluster. Clustering helps in increasing the energy efficiency of WSNs by minimizing the number of communications in the network [2]. In most clustering techniques a cluster head is selected solely on the basis of its energy level. However, there are a few additional essential factors which make a node an ideal candidate to be a cluster head, such as:
- Reachability to nearby nodes
- Quality of connection to base station
- Node reliability [3]

Clustering techniques such as CHUFL[4] and CHEF[5] make use of 2 of these additional factors in selecting a cluster head using fuzzy logic within the WSN. These techniques, through their soft computations, add substantial processing overhead on the WSN. The integration of cloud and WSNs [6] offers the computational power needed. The fuzzy based computation in the proposed algorithm runs on the cloud, thereby eliminating the added processing overhead. The proposed algorithm also caters to all 3 additional factors.

**2. Related Work**

In [1] paper, detailed introduction to wireless sensor networks and their properties has been presented. In [2] paper, distance based Cluster head selection algorithm has been proposed for improving the sensor network life time. In [3] paper, effects of Blackhole attack are measured on the network parameters are presented along with the technique for the detection and prevention of Blackhole attack in WSN. In [4] paper, Cluster Head selection protocol using Fuzzy Logic (CHUFL) has been presented. Cluster Head Election mechanism using Fuzzy Logic protocol (CHEF) is discussed in [5]. In [6] paper, exponential trust based mechanism has been presented to detect malicious nodes. In [12] paper, ACE algorithm evaluates every node's potential one by one for cluster head selection.

**3. Proposed Algorithm: Dynamic Cluster Head Selection Using Fuzzy Logic on Cloud in Wireless Sensor Networks (DCHFC)**

DCHFC operates in 4 phases

- Malicious Node detection [7] and removal in WSN
- Application of fuzzy logic to calculate Potential [12] for each node in Cloud
- Cluster head selection with spatial distribution in Cloud
- Clustering in WSN

*3.1. Malicious Node Detection and Removal*

A trust factor based malicious node detection algorithm [3] is used

> *For each ith node:*
> 1. *Compute Trust Factor of $i^{th}$ node.*
> 2. *$TF_i = 100(x\wedge n)$ where $n_i$ is the no of consecutive data packets dropped.*
> 3. *If $TF_i <= TTF$ (threshold trust factor) value then broadcast $i^{th}$ node as malicious.*

All detected malicious nodes are then removed i.e. not considered for cluster head selection.

*3.2. Application of Fuzzy Logic to Calculate Potential for each Node*

Cloud calculates the Potential value for all nodes, except detected malicious nodes, in the wireless sensor network. Potential is calculated by the following input variables [4]:
- Residual_Energy
- Reachability
- Reception_Power

Where, Residual_Energy is the remaining energy in the node. Reachability is the measure of how accessible the node is to its neighboring nodes, which is defined as [4]

$$r(i) = \frac{1}{N}\left(\sum_{j=1}^{j=N-1} d_{ij}\right) \tag{1}$$

Reception_Power is defined in [4] as Link Quality Indicator/Distance (LQI/Distance) of node from cloud. Distance is the distance between the node and cloud, and LQI has been defined in [8] can be determined by radio chips [9]. Nodes with higher Reception_Power are more likely to become cluster heads [10]. Fuzzy linguistic variables (Low, Medium

High) are assigned to each of these input variables as per their membership functions in [4]. Potential is determined by the Fuzzy rule base table for cluster head selection [4].

### 3.3. Cluster Head Selection with Spatial Distribution

Initially 8-10% of the nodes ($P_{initial}$) with highest Potential are chosen to be cluster heads. Then onwards each remaining node with the highest Potential is taken and its distance D to each of the cluster heads is calculated. If the minimum calculated distance $D_{min} > D_{threshhold}$ the node is selected to be a cluster head, where $D_{threshold}$ is the minimum distance needed to be kept between two cluster heads. If $D_{min} < D_{threshhold}$ the node is rejected i.e. is not made a cluster head, to avoid cluster overlapping.

### 3.4. Clustering in wireless sensor network

Clustering is done in sensor network as per CHUFL rule [4], in which the selected cluster heads broadcast a message in their transmission range. Non-cluster head nodes receive the broadcasted message and select the source of the highest signal strength as its cluster head. On ties, the cluster head with least node ID is chosen. In figure (Fig. 1.) the simulation is carried out assuming that the signal strength is directly proportional to the distance D between nodes.

## 4. Simulation and Analysis

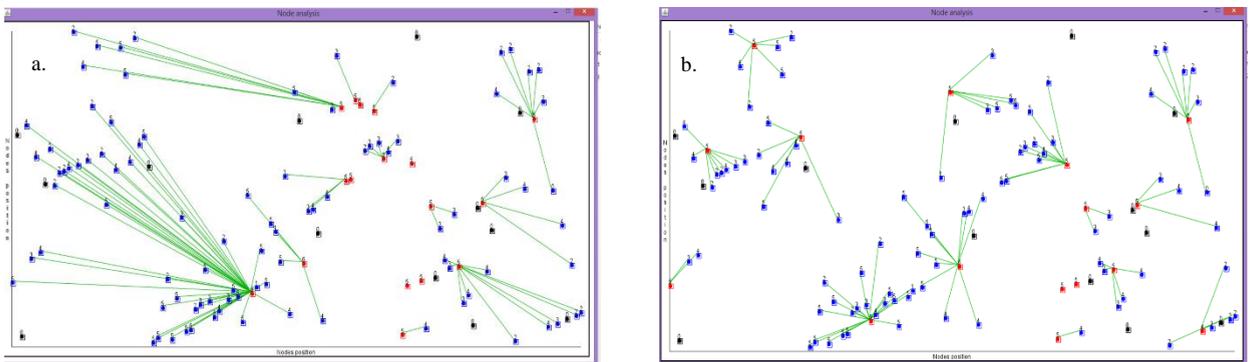

Fig. 1. (a) Cluster formation in CHUFL; (b) Cluster formation in DCHFC

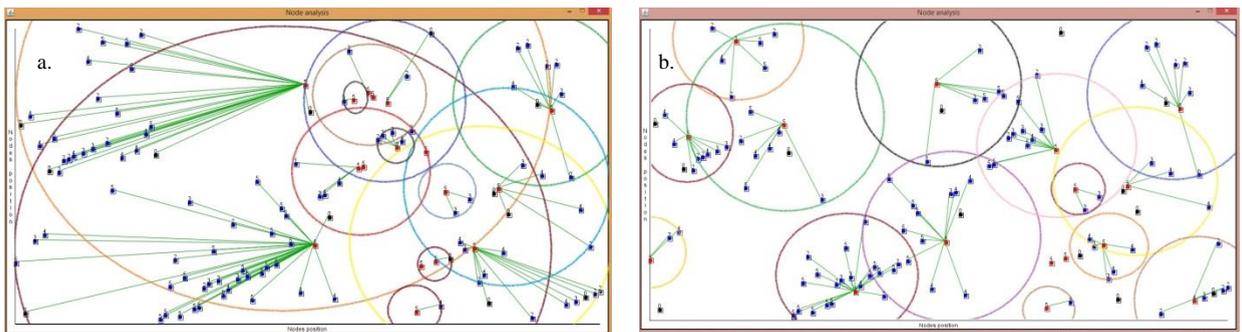

Fig. 2. (a) Cluster overlapping in CHUFL; (b) Cluster overlapping in DCHFC

Fig. 1. (a) shows the cluster formation in WSN using CHUFL[4] by taking 14% nodes as cluster heads and Fig. 1. (b) shows the cluster formation in DCHFC. Table 1. and Table 2. correspond to the simulation of DCHFC and CHUFL respectively. Cluster overlapping is significantly reduced in DCHFC as shown in Fig. 2.

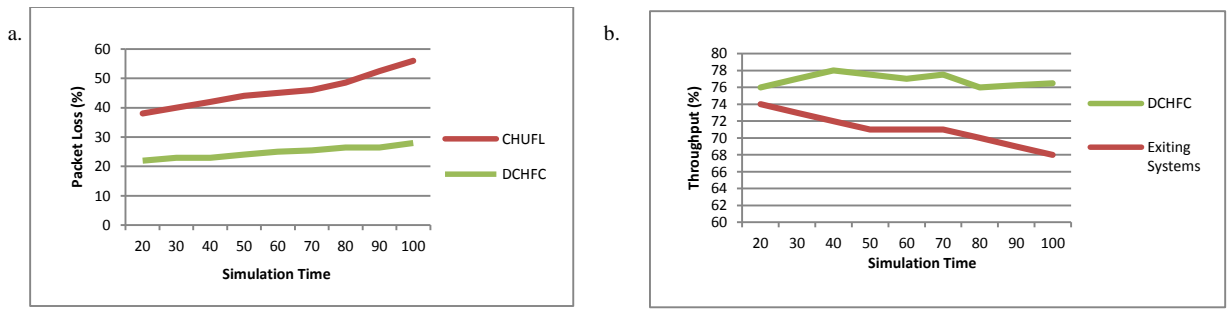

Fig. 3. (a) Packet Loss vs Simulation Time; (b) Throughput vs Simulation Time

Table 1. Simulation of DCHFC

| Simulator | $P_{intitial}$ | Total no. of nodes | No. of cluster heads (red) | No. of malicious nodes (black) | No of clusters | $D_{threshhold}$ |
|---|---|---|---|---|---|---|
| NetBeans[11] | 8% | 122 | 16 | 13 | 14 | 200m |

Table 2. Simulation of CHUFL

| Simulator | Percentage of cluster heads | Total no. of nodes | No. of cluster heads (red) | No. of malicious nodes (black) | No of clusters | $D_{threshhold}$ |
|---|---|---|---|---|---|---|
| NetBeans | 14% | 122 | 17 | 13 | 12 | 200m |

The simulation shows that DCHFC offers spatially distributed cluster heads and creates more clusters to cater to nodes at all distances. Spatially distributed clusters regulate the energy consumption of nodes in the WSN. DCHFC also rejects malicious nodes, resulting in a lower packet loss and higher throughput (Fig. 3.). Simulations show that total residual energy of nodes in DCHFC is higher compared to CHFUL(Fig. 4. a). DCHFC offers a prolonged network lifetime with its nodes surviving longer than CHUFL (Fig. 4. b).

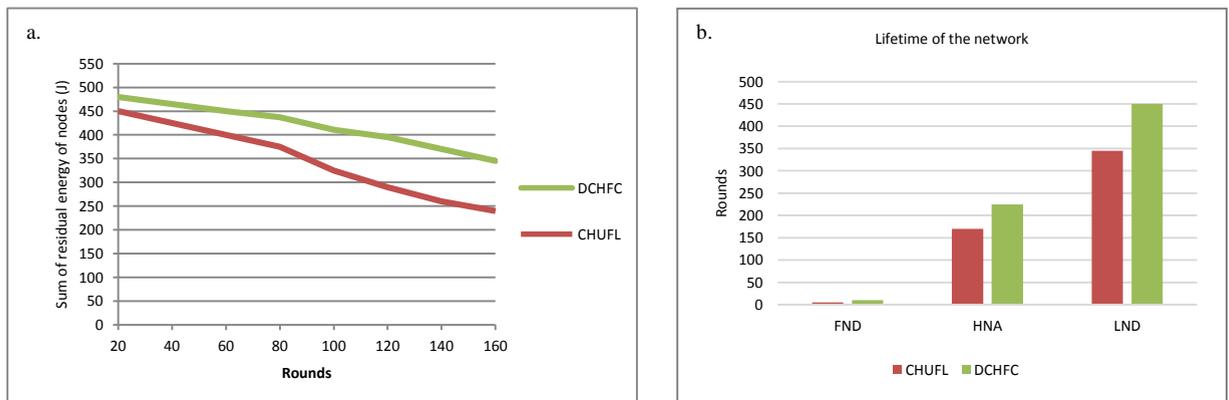

Fig. 4. (a) Total residual energy of network; (b) Time for First Node Dies (FND), Half of the Nodes Alive (HNA) and Last Node Dies (LND)

## 5. CONCLUSION

Protocols for Wireless Sensor Networks need to be energy efficient to prolong network lifetime. To achieve enhancement in network lifetime and reliable and efficient communication in WSNs, deploying a good clustering technique is essential. The proposed algorithm (DCHFC) for clustering offers spatially distributed cluster heads selection, which reduces cluster overlapping. DCHFC detects malicious nodes and never allows them to be the cluster head; thereby DCHFC increases the reliability and makes sensor networks fault tolerant. Cluster head selection by DCHFC is based on potential value calculated using fuzzy. The spatial distribution of clusters prevents overlapping. Having all these features DCHFC proves to be the better choice for energy consumption and enhancement of network lifetime.